\documentclass[twocolumn,aps,amssymb,floatfix,prl,showpacs]{revtex4}
\usepackage{amsmath}
\usepackage{amsfonts}
\usepackage{graphicx}

\begin{document}

\title{Control of Coulomb blockade in a mesoscopic Josephson
junction using single electron tunneling}

\author{J. Hassel$^1$, J. Delahaye$^2$, H. Sepp\"a$^1$, and P. Hakonen$^2$}
\address{$^1$VTT Information Technology, Microsensing, P.O. Box 1207,
FIN-02044 VTT\\
$^2$Low Temperature Laboratory, Helsinki University of Technology,
 P.O. Box 2200, FIN-02015 HUT}

\begin{abstract}
We study a circuit where a mesoscopic Josephson junction (JJ) is
embedded in an environment consisting of a large bias resistor and
a normal metal - superconductor tunnel junction (NIS).
The effective Coulomb blockade of the JJ can be controlled by the tunneling
current through the NIS junction leading to transistor-like
characteristics. We show using phase correlation theory and
numerical simulations that substantial current gain with low
current noise ($i_{n}\lesssim 1$ fA/$\sqrt{\text{Hz}}$) and
noise temperature ($\lesssim $ 0.1 K) can be achieved. Good agreement
between our numerical simulations and experimental results is
obtained.

\end{abstract}

 \bigskip

\maketitle

Mesoscopic Josephson junctions (JJ) display interesting phenomena
owing to the conjugate nature of phase and charge \cite{Tinkham}.
Coulomb effects cause delocalization of the macroscopic phase
variable across the tunnel junction, which leads to the formation
of energy bands \cite{ALZ}. As the band width grows rapidly with
the band index, a mesoscopic JJ makes it possible to construct
novel devices where the operation is based on controlling the
transitions between energy levels of the junction, thereby
controlling the effective Coulomb blockade of the device
\cite{sep1,has1}. The control is made using a small tunnel current
of single electrons. Thus, the device distinguishes from the
ordinary Coulomb blockade devices, like single electron
transistors (SETs), where the current is adjusted by an external,
capacitively-coupled voltage \cite{review}.

We have investigated the circuit where a mesoscopic Josephson
junction (JJ) is embedded in an environment consisting of a large
bias resistor and a normal metal - superconductor tunnel junction
(NIS). The JJ is biased in the regime where the system becomes a
two-level system with two distinct Coulomb blockade strengths. The
effective Coulomb blockade of the JJ is controlled by the
tunneling current through the NIS junction, which leads to
transistor-like characteristics. This device, called the Bloch
oscillating transistor (BOT) \cite{has1}, provides a low-noise
current amplifier whose input impedance level makes it an
intermediate device between the ultimate low temperature
amplifiers, the SQUID and the SET \cite{science}.

In this Article, we present experimental results on the basic
properties of BOTs, and compare them with computer simulations
based on time dependent phase correlation theory for electron and
Cooper pair tunneling. We show that this provides a way to model
the devices quantitatively. Noise properties of the devices are
discussed, the conclusion being that ultra low-noise current
amplifiers (current noise $i_{n}\lesssim 1$ fA/$\sqrt{\text{Hz}}$
referred to input)
can be built on the basis of controlled JJs.
The simulated results are shown to be
in a good agreement with our experimental findings.

Schematically our device is shown in Fig. 1 (left frame). A
Josephson junction (JJ) connects a superconducting island to the
emitter electrode (E). A Normal-Insulator-Superconductor (NIS)
junction connects the normal base electrode (B) to the island.
Furthermore, a high-impedance, thin film resistor $R_{C}$ connects
the island to the collector electrode (C). When the isolation
resistance $R_{C}\ggg R_{Q}=h/4e^{2}$, charge fluctuations on the
JJ are small and the dynamics of the junction consists purely of
Bloch reflections at the Brillouin zone boundary, interrupted by
occasional Zener tunneling between the bands \cite{Zener}.

 \begin{figure}

    \includegraphics[width=7.5cm]{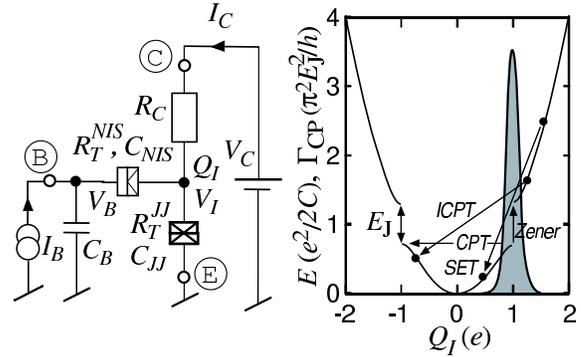}

    \caption{Schematics of the circuit (left frame) for
    controlling the Coulomb blockade of Cooper pairs. The injector of single electron
    current, the superconductor-normal junction, is denoted by NIS while the Josephson
    junction is coined by JJ. Symbols for different biasing voltages and the
    resulting currents are also given in the figure. $Q_I$ denotes the island
    charge that is governed by Eq. (1). See the
    caption of Table I for other parameters.
    The right frame illustrates the energy vs. charge diagram
    used in the simulations. The distribution at the Brillouin zone boundary
    illustrates the charge fluctuations due to the resistive
    environment.
    Arrows indicate possible tunneling processes in the
    BOT: CPT - Cooper pair tunneling, ICPT - inelastic Cooper
    pair tunneling, SET - single electron tunneling, Zener -
    tunneling up to higher energy band. $\Gamma_{CP}$ denotes the tunneling rate for both CPT and
    ICPT.
     }\label{Scheme}

    \end{figure}

Fundamentally, the operation of a BOT is understood on the basis
of the band structure of a Josephson junction. The energy versus
(quasi)charge $Q_{I}$ of a JJ is shown with a thick uniform curve
in the right frame of Fig. 1. When the collector is voltage biased
by $V_C$, the "island" charge $Q_I$ tends to relax through $R_{C}$
towards the value
$Q_{I}=(C_{JJ}+C_{NIS})V_{C}-C_{NIS}V_{B}\approx C_{\Sigma }V_{C}$,
where $C_{\Sigma }=C_{JJ}+C_{NIS}$.
At $Q_{I}=e$, Cooper pair (CP) tunneling returns the system back
to the state $Q_{I}=-e$. Repetition of this Bloch oscillation
cycle produces a net current from the collector to the emitter.
The competing process for the CP tunneling at $Q_{I}=e$ is Zener
tunneling up to the second energy band, where Bloch oscillations
do not occur and the current flow stops (Coulomb blockade voltage
$\partial E/\partial Q_I > V_C$). From the second band the
system may return back to the first one by single-electron
tunneling through the base junction. Thus, as a result of a single
electron tunneling event through the NIS junction, a sequence of
Cooper pair pulses is injected through the JJ, which leads to a current gain
equaling the average number of electrons in the CP sequence. This
kind of action is below referred to as regular BOT operation.

The above picture is, however, an overidealization in most
experimental cases. In a real situation all input electrons do not
produce output current pulses, but rather cause intraband
transitions. It is also hard to fabricate large enough thin film
resistances to achieve coherent Bloch oscillations. Qualitatively,
the picture of the two-level system still works, but the measured
current gains are clearly less than the calculated values
\cite{PhysicaE}. When $R_C$ becomes relatively small, $\sim 10
R_Q$, charge fluctuations grow and inelastic Cooper pair tunneling
may take place clearly before the Brillouin zone boundary. This
makes the dynamics more complicated and our present experiments
and simulations are aimed at clarifying this question.

\begin{table}
\begin{tabular}{|c|c|c|c|c|c|c|}
   \hline  & $R_T^{JJ}$ & $R_T^{NIS}$&
   $R_{C}$ & $C_{NIS}$ (fF) & $C_{JJ}$ (fF) & $E_J^{min}$ / $E_J^{max}$ \\
   \hline 1 & 8.1 & 27.3 & 23 & 0.28 & 0.95 & 22 / 78 \\
   \hline 2 & 7.8 & 5.8  & 50 & 0.6 & 1.0 & 83 / 83  \\
\hline
\end{tabular}
  \caption{Parameters for our samples. Tunneling resistances
  (in k$ \Omega )$ of the Josephson and NIS
  junctions are given by $R_T^{JJ}$ and
  $R_T^{NIS}$, respectively. $R_{C}$ denotes the
  environmental impedance of the Josephson junctions.
  The division of capacitance between the two junctions, $C_{NIS}$ and $C_{JJ}$,
  are estimated on the basis
  of the measured resistances and geometrical dimensions. The last column indicates the
  minimum $E_J^{min}$ and maximum $E_J^{max}$ values of the Josephson
  energy in $\mu$eV.}

\end{table}

In our computational model the island charge as function of time
is obtained by integrating the equation%
\begin{equation}
\frac{dQ_{I}}{dt}=\frac{V_{C}-V_{I}}{R_{C}}-\left(
\frac{dQ_{I}}{dt}\right) _{NIS}-\left(  \frac{dQ_{I}}{dt}\right)
_{JJ}, \label{charge}
\end{equation}

\noindent where the first term represents the charge relaxation
through the collector resistor, and the last two terms represent
tunneling current in the NIS and JJ, respectively.  $\left(
dQ_I/dt\right) _{NIS}$ contains only quasiparticle current while
$\left( dQ_I/dt\right) _{JJ}$ includes both quasiparticle and
Cooper pair tunneling. The base junction is current biased, as in
the experiments, but the gate capacitance $C_B \sim 1$ pF (see
Fig. 1) converts it effectively into voltage bias $V_B$
\cite{NOTE}.

To integrate Eq. (\ref{charge}) in the presence of finite
Josephson coupling and electromagnetic environment, we compute the
Cooper pair tunneling through the JJ ($\Gamma_{CP}$) and
quasiparticle tunneling through both junctions
($\Gamma_{QP}^{JJ}$, $\Gamma_{QP}^{NIS}$) using the
phase-fluctuation $P\left( E\right) $-theory \cite{dev1,ing1}. As
a modification to the standard theory, we use time dependent
voltages $V_{I}$ and $V_{I}-V_{B}$ across the JJ and NIS
junctions, respectively. $V_I$ is given by
\begin{equation}
V_{I}=\left( C_{NIS}/C_{\Sigma }\right) V_{B}+Q_{I}/C_{\Sigma }.\label{VI}
\end{equation}

\noindent A basically similar modelling approach has been employed
by Kuzmin \textit{et al.} \cite{Kuzmin96} when investigating the
role of Zener tunneling on the IV-curves of ultra small Josephson
junctions. The difference is that we include the effect of quantum
fluctuations in our model on top of the thermal noise which was
employed in Ref. \cite{Zaikin92} only.

According to the $P\left( E\right) $-theory, tunneling does not
happen strictly at $Q_{I}=e$, but is rather represented with a
finite distribution, which is schematically shown in Fig.
\ref{Scheme}. Within this picture, one can still use the concept
of band structure, the interpretation being that values
$Q_{I}\lesssim e$ correspond to the first band, and $Q_{I}\gtrsim
e$ to the second band. The band gaps, where Zener tunneling takes
place, are now reflected in the probabilities at which the
junction may pass from the lower to the higher band (see below).

The most critical assumption is made while computing the Cooper
pair tunneling rate, for which the lowest order theory is valid
when $E_{J}P\left( 2eV\right) \ll 1$ \cite{ing1}. For the samples
analyzed in this paper, $\max \left( P\left( 2eV\right) \right)
\approx 0.3/E_{C}$ and $\max \left( E_{J}\right) \approx
1.7E_{C}$, so that second order effects may be expected to be small. The Cooper pair
tunneling rate can then be computed as
\begin{equation}
\Gamma_{CP}\left(  V_{I}\right)  =\frac{\pi
E_{J}^{2}}{2\hbar}P\left(
2eV_{I}\right)  ,\label{GCP}%
\end{equation}
where the function $P\left(E\right)  $ is defined as%
\begin{equation}
P\left(  E\right)  =\frac{1}{2\pi\hbar}\int\limits_{-\infty}^{\infty}%
dt\exp\left(  J\left(  t\right)  +\frac{i}{\hbar}Et\right)  .\label{pe}%
\end{equation}

\noindent The phase correlation function $J(t) =\langle[
\varphi(t) -\varphi(0)]\varphi(0)\rangle$ which takes into account
the fluctuations of the  phase $\varphi(t) $ on the junction.
$J(t)$ is calculable from the real part of the environmental
impedance \cite{ing1}. Here we have also made the assumption that
the effect of the environment is exclusively due to the collector
resistance \cite{note}, for which the $P(E)$ function was calculated numerically.

In the limit $R_{C}/R_{Q}\gg E_{C}/kT$, the CP tunneling rate as
given by Eq. \ref{GCP} becomes a Gaussian distribution centered
around $V_{I}=e/C_{\Sigma }$ \cite{ing1}. If furthermore $E_{C}/kT
\gg 1$, the distribution narrows to a delta spike $\Gamma
_{CP}=\left( \pi E_{J}^{2}/2\hbar \right) \delta \left(
2eV-4E_{c}\right) $ which equals the equation obtained from the
band model by neglecting the effect of the environment
\cite{sch1}. Hence, the basic features of the band model are
embedded in the peaked tunneling probabilities. In a proper band
model, however, the capacitance of the JJ entering Eq.
(\ref{charge}) would be nonlinear and given by $(d^2E/dQ^2)^{-1}$.
In our model, we take the capacitance as constant, which is valid
in the limit $E_J \longrightarrow 0$ only.

Our sample parameters are given in Table I. Details of sample
manufacturing and experimental techniques can be found in Ref.
\cite{science}. Measured and computed $I_{C}V_{C}$-curves for
sample I with various $E_J$ are shown in Fig. 2, each at several
values of $I_B$. When $I_B=0$, a weak Coulomb blockade is visible
at zero bias. The peak in current at non-zero voltages reflects
Cooper pair tunneling processes, which are enhanced by
single-electron current through the base. The current gain $\beta$
is found to be maximized in the region with negative slope:
$\beta_{max}= 3.2$ at $V < 0$ ($I_B=3.3$ nA) and $\beta_{max}=
3.0$ at $V > 0$ ($I_B=2.5$ nA). From our simulations we get
$\beta_{max}= 2.8$ ($I_B=1.7$ nA) and 2.5 ($I_B=1.7$ nA),
respectively. The relatively small maximum gains are caused by
large current fluctuations owing to $R_C=23$ k$\Omega$. Fig. 3
displays $I_{C}V_{C}$-curves for sample II with $E_J/E_C=1.7$.
Both the measured data \cite{science} and the simulated
$I_{C}V_{C}$-curves display hysteretic behavior. The computed
curves in the regular BOT regime are seen to display growing
hysteresis with increasing base current, in accordance with the
measured data. Especially with large $E_J$, the current peak is at
slightly lower voltages in the computed curves than in the
measured data indicating that the computed probability of Zener
tunneling is larger than in reality. This is a sign that our
simplified picture of the energy band structure fails for large
$E_J$.
%The reason is probably the
%simplified picture of the energy band structure, which fails for
%large $E_J$.

   \begin{figure}

    \includegraphics[width=8.5cm]{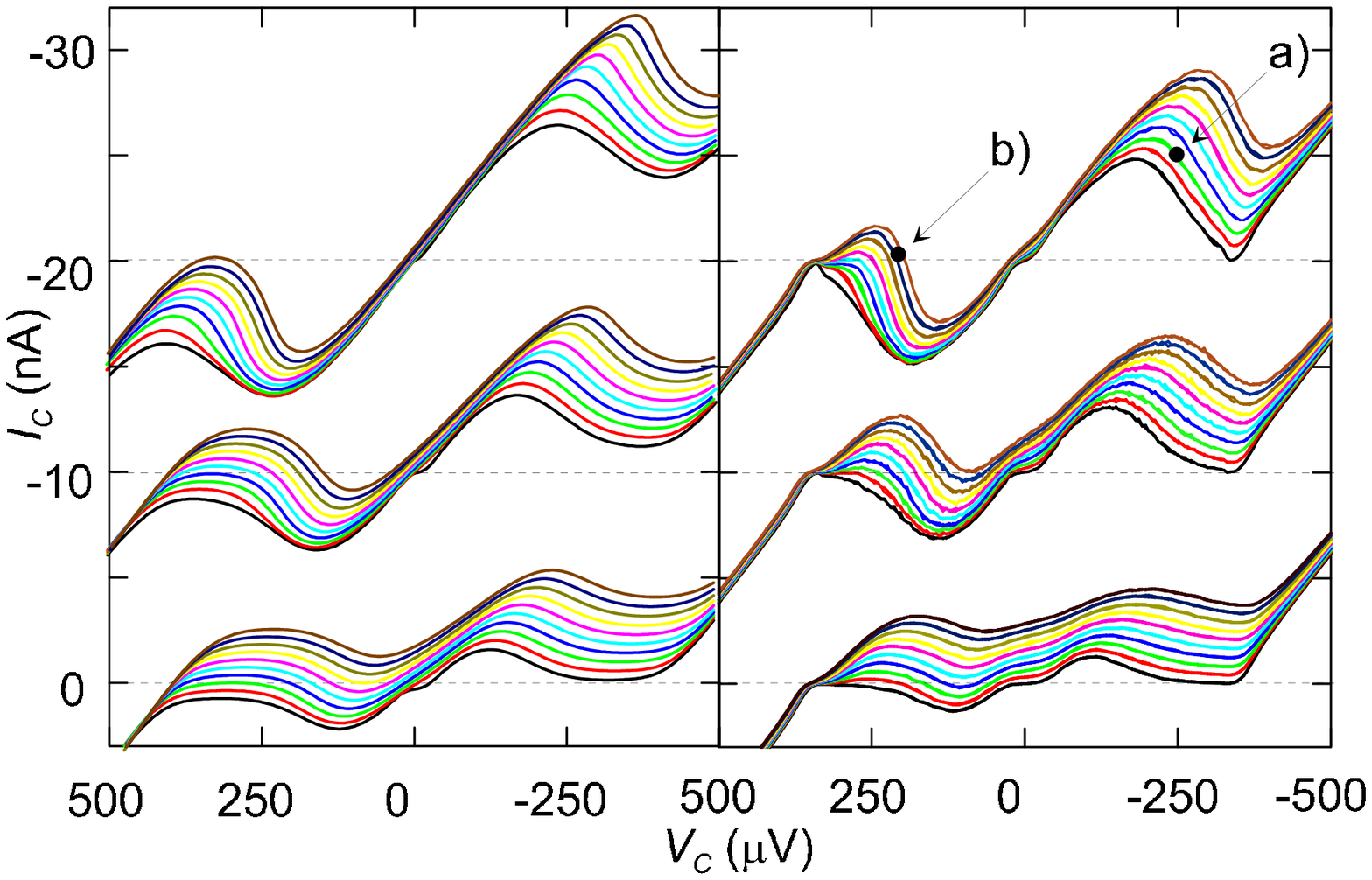}

    \caption{Sets of $I_C V_C$-curves measured (left) and computed (right)
    at $T=90$ mK on sample 1.  Josephson coupling has been varied with $E_J/E_C=1.2$
    at the topmost, $E_J/E_C=0.7$ at the middle and $E_J/E_C=0.35$ at the lowest set.
    In each set the base currents $I_B$=0, 0.4, 0.8,
    1.2, 1.6, 2.0, 2.4, 2.8, 3.2, and 3.6 nA (in order from bottom to top).
    The topmost set is offset by -20 nA and middle set by
    -10 nA for clarity.}\label{IVsample1}
    \end{figure}

   \begin{figure}

    \includegraphics[width=7cm]{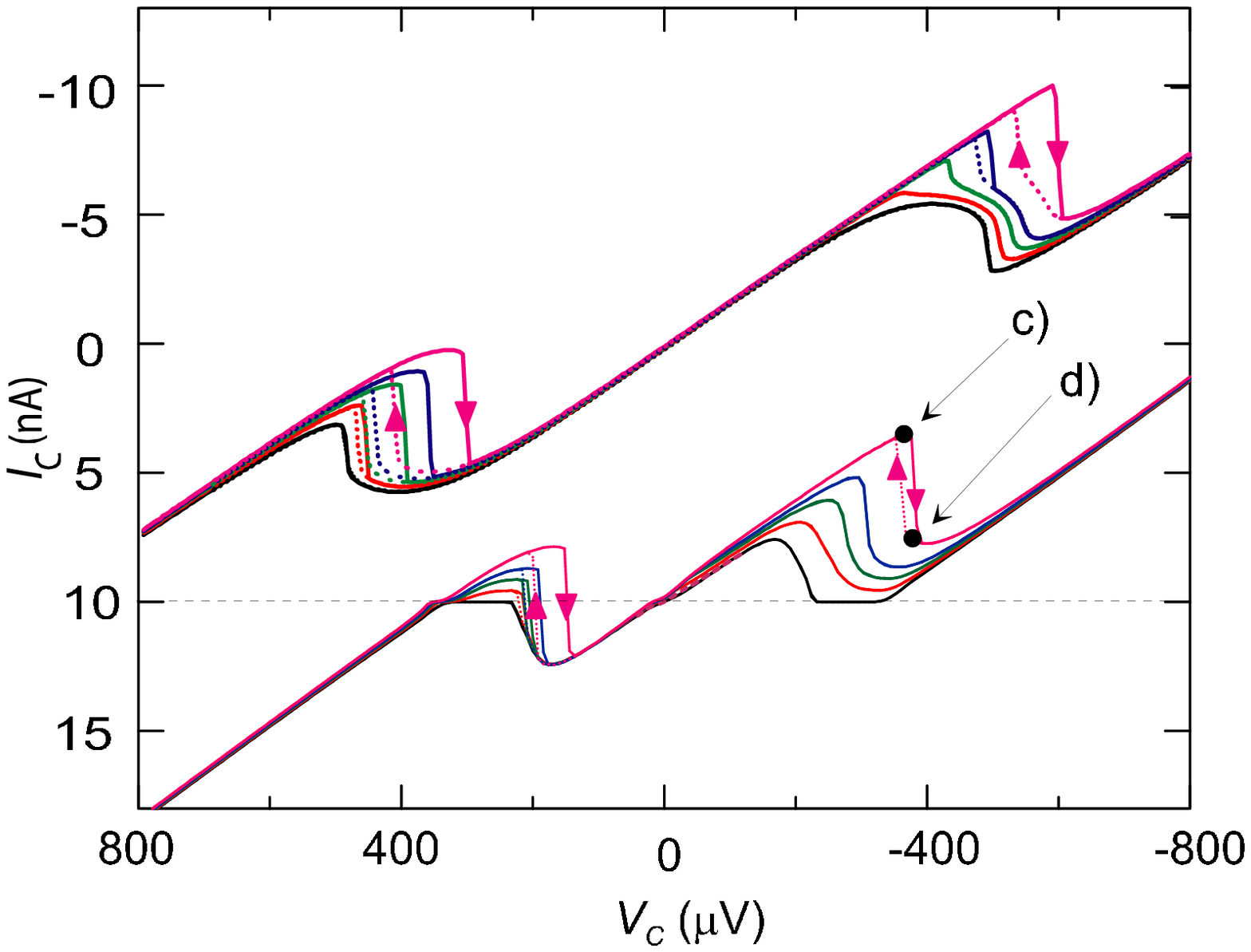}

    \caption{$I_C V_C$-curves
    for sample 2 with $E_J/E_C=1.7$ at base currents $I_B$=0, 0.4, 0.8,
    1.2, and 2.0 nA. The upper set represents data measured at $T=100$ mK \cite{science}.
    The lower set is the result of simulation using the experimental parameters. It 
    is offset by 10 nA for clarity.}\label{IVsample2}

   \end{figure}

The nonsymmetrical nature of the curves in Fig. 2 suggests that
the mechanism for current gain is different at opposite biasing
polarities. This is illustrated in the time traces of $Q_{I}$ in
Figs. 4a and 4b. In the regular biasing case ($V_C <0$, Fig. 4a)
single-electron tunneling is seen to drive the system downwards
from the upper band and, thus, it tries to restore Bloch
oscillations, \textit{i.e.} the dynamics is essentially similar to
the original BOT operation even though incoherent tunneling
phenomena modifies it strongly. When $V_C
>0$, on the contrary, $I_B$ tends to drive the system into the
second band, which leads to a suppression of 2e-oscillations as
seen in Fig. 4b. The recovery of 2e-oscillations now takes place
through incoherent 2e-tunneling, the tunneling probability of
which is determined by the ''tail'' of $P\left( E\right)
$-function.

Time domain plots of $Q_I$ for the two hysteretic $I_C$-branches
(see Fig. 3) are shown in Figs. 4c and 4d. On the upper branch
(Fig. 4c) the dynamics is again regular BOT dynamics, essentially
similar to that of Fig. 4a. Now it is obvious that the
system almost never relaxes to its stationary state, since the
inverse of $R_{C}C_{\Sigma}$-time constant is small compared to
the tunneling rate $\Gamma_{NIS}$. Single electron tunneling still
clearly enhances Cooper pair tunneling. At the lower branch, the
system seldom returns to the lowest band, and the base current
$I_B$ mostly consists of tunneling events causing intraband
transitions only. Therefore, Cooper pair tunneling is less likely
and the collector current remains small. Two stable solutions can
coexist since $I_{B} \propto (V_{B}-V_{JJ})$. Now, due to the
current biased base electrode, $V_{B}$ and $V_{JJ}$ can each
dynamically assume two different average values, while their
difference remains the same leading to equal base currents in both
cases.

   \begin{figure}

    \includegraphics[width=7cm]{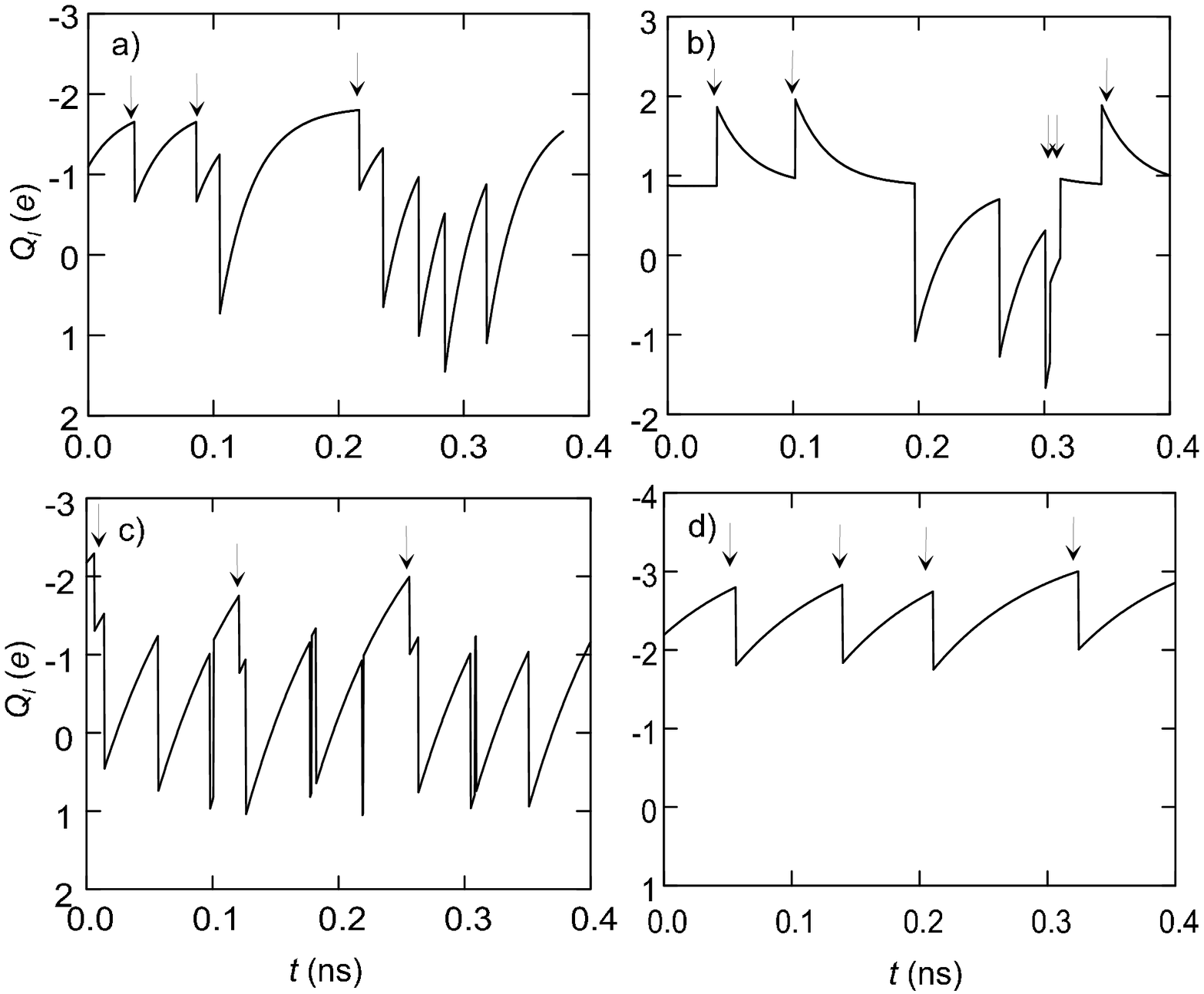}

    \caption{Time traces of charge $Q_{I}$ at the operating points denoted by
             solid circles in Figs. 2 and 3: a) regular BOT operation,
             b) inverted BOT operation, c) regular BOT operation in the
             hysteretic region, and d) operation in the hysteretic region
             with only intraband transitions induced by $I_B$. The arrows indicate
             single electron tunneling events.}\label{TimeDomain}

    \end{figure}

In our simulations, the transition from non--hysteretic to
hysteretic behavior can be crossed by increasing $E_J$. This is because
increasing $E_J$ enhances the stability of the type of
solution presented in Fig. 4(c) by increasing the rate
$\Gamma_{CP}$ according to Eq. (\ref{GCP}). Near the hysteresis
point, the ratio of base tunneling currents, $I_{B_1}/I_{B_2}$,
made of contributions causing either interband transitions
($I_{B_1}$) or intraband transitions ($I_{B_2}$) is found to vary
steeply, for example as a function of $I_B$. Consequently, a small
change in the input current induces a large change in the output
current, because of the conversion of base current between the
types $I_{B_1}$ and $I_{B_2}$. Large gain may then be obtained with
ultra low noise. The output noise current is independent of 
source impedance if $C_{B}\gg C_{\Sigma }$. In this case
the noise temperature can be written 

\begin{equation}\label{Tn}
  T_N=Z_{in} \frac{i_{n_{out}}^2}{\beta^2} ,
\end{equation}

\noindent where $Z_{in}$ denotes the input impedance and
$i_{n_{out}}/\beta$ is the output current noise converted to the
input. Near the hysteresis point, $\beta$ may grow without limit,
which makes it possible to reduce $T_N\longrightarrow 0$. Note,
however, that this takes place at the cost of input impedance
$\sim \beta R_C$ and maximum acceptable input signal amplitude.
Experimentally, we have found a sample with white noise of $i_{n}
=$ 10 fA$/\sqrt{\text{Hz}}$ referred to the input. This implies
a noise temperature $T_N$ smaller by a factor of five
compared to the shot noise
approximation $i_{n}^2=2 e I_B$. The result was measured 
on a sample having a Josephson coupling energy of order
$4E_{C}$, \textit{i.e.} slightly out of
the validity range of our computational model. In our simulations,
current noise values of order 1 fA$/\sqrt{\text{Hz}}$ and noise
temperatures below 0.1K have been reached using experimentally
realizable parameters.

As mentioned in the beginning, our analysis is based on
perturbation theory in $E_J/E_C$ and the results are strictly valid in the
limit $E_J\longrightarrow 0$. Nevertheless, we believe that
our simulations are principally valid at
values of $E_J \simeq 1$. More serious problems occur when the width
of the lowest band becomes exponentially narrow as for samples
with $E_{J}/E_{C} \sim 4$ \cite{science}. In this case, the
dynamics is dominated by transitions between higher bands, rather
than the two lowest ones. As long as there are two major bands
involved allowing to use a two band approximation, our simulated
results will be qualitatively correct.

In summary, we have shown that novel devices can be constructed
using interlevel transitions, driven by single electron tunneling.
The single electron current can be used either to drive the JJ in
to the blockade state, or out from the blockade. Both methods are
seen to yield substantial current gain, though at slightly
different values of bias voltage. Comparison of our experimental
and theoretical results show that time-dependent $P(E)$-theory can
be employed quite successfully to model the behavior of such
devices. The essential features of BOT dynamics were found to be recovered
even in the presence of relatively strong
incoherent Cooper pair tunneling and the JJ could still be
understood as a two-level system, whose switching is controlled by
single charge tunneling. Furthermore, our simulations indicate
that intraband transitions play a significant role in these three
terminal devices. The devices may give substantial gain by
conversion of base current between interlevel and intraband types.
This leads to noise powers that are substantially less than the
values obtained from the input shot noise approximation.

We acknowledge fruitful discussions with R. Lindell, T. Heikkil\"a, F.
Hekking, G.-L. Ingold, A. Niskanen, M. Paalanen, M.
Sillanp\"a\"a, M. Kiviranta and A. Zaikin. This work was supported by the
Academy of Finland and by the Large Scale Installation Program
ULTI-3 of the European Union.


\begin{thebibliography}{99}
\bibitem{NOTE} $V_B$ was calculated using the equation
$\frac{dV_{B}}{dt}=\frac{1}{C_{B}}\left(  I_{B}+\left(
\frac{dQ_{I}}{dt}\right)  _{NIS}\right).$ It was assumed that
$C_{B}\gg C_{\Sigma}$.
\end{thebibliography}
\end{document}